\documentclass[sigconf]{acmart}
\usepackage{enumitem}
\usepackage{array,hhline}
\usepackage{graphicx}
\usepackage[ruled,linesnumbered]{algorithm2e}
\usepackage{microtype}
\usepackage{lipsum}
\usepackage{supertabular}
\usepackage{wrapfig}
\usepackage{multirow}
\usepackage{array,hhline}
\usepackage{graphicx}
\usepackage{framed}
\usepackage{setspace}
\usepackage{xcolor}
\usepackage{caption}
\usepackage{enumitem}
\usepackage{supertabular}
\usepackage{fontawesome}
\usepackage{dsfont}
\usepackage[T1]{fontenc}
\usepackage{microtype}      
\usepackage{xcolor}         
\usepackage{xspace}
\usepackage{url}
\usepackage{mathtools}
\usepackage{booktabs}
\usepackage[medium, compact]{titlesec}

\copyrightyear{2024}
\acmYear{2024}
\setcopyright{rightsretained}
\acmConference[KDD '24]{Proceedings of the 30th ACM SIGKDD Conference on Knowledge Discovery and Data Mining}{August 25--29, 2024}{Barcelona, Spain}
\acmBooktitle{Proceedings of the 30th ACM SIGKDD Conference on Knowledge Discovery and Data Mining (KDD '24), August 25--29, 2024, Barcelona, Spain}\acmDOI{10.1145/3637528.3671511}
\acmISBN{979-8-4007-0490-1/24/08}

\makeatletter
\gdef\@copyrightpermission{
  \begin{minipage}{0.3\columnwidth}
   \href{https://creativecommons.org/licenses/by/4.0/}{\includegraphics[width=0.90\textwidth]{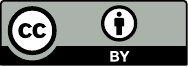}}
  \end{minipage}\hfill
  \begin{minipage}{0.7\columnwidth}
   \href{https://creativecommons.org/licenses/by/4.0/}{This work is licensed under a Creative Commons Attribution International 4.0 License.}
  \end{minipage}
  \vspace{5pt}
}
\makeatother

\begin{document}

\title{MMBee: Live Streaming Gift-Sending Recommendations via Multi-Modal Fusion and Behaviour Expansion}

\author{Jiaxin Deng}
\authornote{Equal contribution.}
\authornote{
This work is done when Jiaxin Deng is an intern at KuaiShou.}
\affiliation{%
  \institution{Institute of Automation}
  \country{}
}
\affiliation{%
  \institution{School of Artificial Intelligence, University of Chinese Academy of Sciences}
  \country{Beijing, China}
}
 \email{dengjiaxin2022@ia.ac.cn}
\author{Shiyao Wang}
\authornotemark[1]
\affiliation{%
  \institution{KuaiShou Inc.}
   \country{Beijing, China}
   }
   \email{wangshiyao08@kuaishou.com}
\author{Yuchen Wang}
\affiliation{%
  \institution{KuaiShou Inc.}
   \country{Beijing, China}
   }
   \email{wangyuchen11@kuaishou.com}
\author{Jiansong Qi}
\affiliation{%
  \institution{KuaiShou Inc.}
   \country{Beijing, China}
   }
   \email{qijiansong@kuaishou.com}
\author{Liqin Zhao}
\affiliation{%
  \institution{KuaiShou Inc.}
   \country{Beijing, China}
   }
   \email{zhaoliqin@kuaishou.com}
\author{Guorui Zhou}
\authornote{Corresponding author.}
\affiliation{%
  \institution{KuaiShou Inc.}
   \country{Beijing, China}
   }
   \email{zhouguorui@kuaishou.com}
\author{Gaofeng Meng}
\affiliation{%
  \institution{Institute of Automation}
   \country{Beijing, China}
   }
   \email{gfmeng@nlpr.ia.ac.cn}


\newtheorem{dfn}{Definition}
\renewcommand{\shortauthors}{Jiaxin Deng et al.}

\begin{abstract}
  Live streaming services are becoming increasingly popular due to real-time interactions and entertainment. Viewers can chat and send comments or virtual gifts to express their preferences for the streamers. Accurately modeling the gifting interaction not only enhances users' experience but also increases streamers' revenue. Previous studies on live streaming gifting prediction treat this task as a conventional recommendation problem, and model users' preferences using categorical data and observed historical behaviors. However, it is challenging to precisely describe the \textit{\textbf{real-time content changes}} in live streaming using limited categorical information. Moreover, due to the \textit{\textbf{sparsity of gifting behaviors}}, capturing the preferences and intentions of users is quite difficult. In this work, we propose \textbf{MMBee} based on real-time \underline{M}ulti-\underline{M}odal Fusion and \underline{Be}haviour \underline{E}xpansion to address these issues. Specifically, we first present a Multi-modal Fusion Module with Learnable Query (MFQ) to perceive the dynamic content of streaming segments and process complex multi-modal interactions, including images, text comments and speech. To alleviate the sparsity issue of gifting behaviors, we present a novel Graph-guided Interest Expansion (GIE) approach that learns both user and streamer representations on large-scale gifting graphs with multi-modal attributes. It consists of two main parts: graph node representations pre-training and metapath-based behavior expansion, all of which help model jump out of the specific historical gifting behaviors for exploration and largely enrich the behavior representations. Comprehensive experiment results show that MMBee achieves significant performance improvements on both public datasets and Kuaishou real-world streaming datasets and the effectiveness has been further validated through online A/B experiments. MMBee has been deployed and is serving hundreds of millions of users at Kuaishou.
\end{abstract}



\maketitle

\section{Introduction}
\vspace{-0.1cm}

\begin{figure}[h]
\centering
\vspace{-0.3cm}
\includegraphics[width=.43\textwidth]{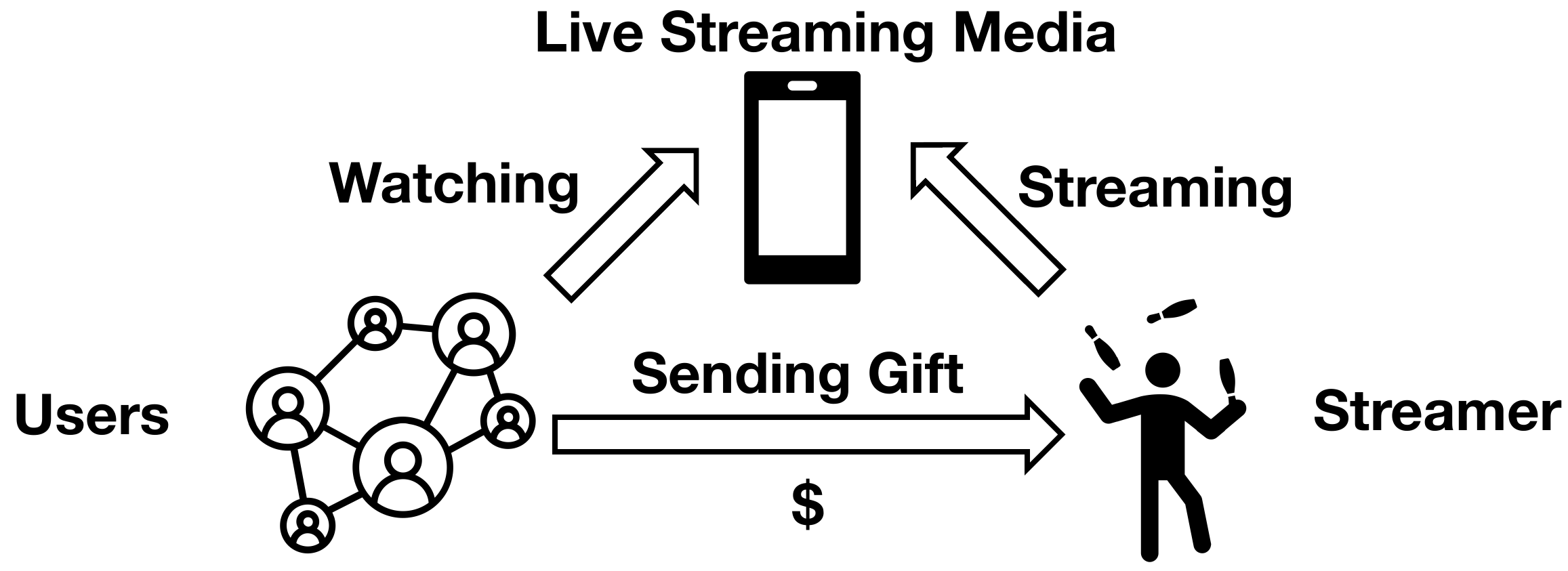}
\caption{\textbf{Example of the live streaming gifting scenario with the interactions among users and streamers.}}
\label{fig1}
\vspace{-0.5cm}
\end{figure}
Due to the rapid development of mobile device hardware and the Internet, live streaming has become a prevalent social service for people's daily lives. As one of the most popular live streaming platforms in China, \textit{Kuaishou} has reached 386.6 million daily active users and the revenue generated by the live streaming business reached RMB 9.7 billion as of the third quarter of 2023, which heavily relies on Kuaishou's continuous optimization of the live streaming ecosystem and improvement of the recommendation system. As shown in Figure \ref{fig1}, on live streaming platforms, content creators can share their produced video content with users in real-time, and users can interact with streamers and peers through live comments or discussions. They can even send virtual gifts to their favorite streamers, which is one of the main sources of revenue for the live-streaming business. Therefore, the task of live streaming gifting prediction is vital not only for enhancing user experience and streamer revenue but also for increasing the business effectiveness of the platform.

Recent years have witnessed several relevant methods for recommendation \cite{lai2023learning,yu2021leveraging, yu2020social, rappaz2021recommendation, tu2018earning} and gifting prediction \cite{lai2020live, xi2023multimodal} in live streaming. For example, MARS \cite{lai2020live} introduces a two-stage recommendation approach applied in the Multi-Stream Party scenario, aiming to maximize reward earnings while optimizing user personal experience at the same time. However, this approach ignores the close connection between users’ gifting behavior and the rapidly changing live content in the living room. To address this issue, MTA \cite{xi2023multimodal} designs a novel orthogonal module that fully utilizes the multi-modal features in live streaming. However, MTA treats the gift prediction as a time series prediction problem which does not consider users' personalization. 
Although typical behavior-based methods like SIM \cite{pi2020search} can achieve personalized recommendations for gifting prediction, they may face the challenge of behavior sparsity in the context of live streaming. According to \cite{grbovic2018real}, DNN-based methods typically require a minimum of 5-10 historical behavior sequences to learn meaningful representations for modeling user interests. However, the average length of user's gifting behavior is as low as 0.3 anchors in our scenario. Therefore, gifting prediction requires a comprehensive consideration that combines user personalization under sparse behaviors and real-time content modeling to achieve optimal recommendation effectiveness.

To address these challenges, we propose \textbf{MMBee}: an efficient live streaming gifting prediction method based on real-time \underline{M}ulti-\underline{M}odal Fusion and \underline{Be}haviour \underline{E}xpansion. Specifically, we first design a Multi-modal Fusion Module with Learnable Query (MFQ). It helps the model to perceive the \textit{\textbf{real-time content changes}} in live streaming through processing the complex visual frames, comments and audio in each streaming segment. In addition, aiming to address the \textit{\textbf{sparsity problem}} in gifting prediction, we propose a novel Graph-guided Interest Expansion (GIE) approach. We first construct large-scale gifting graphs based on the history of gifting interactions. Then a graph pre-training scheme via contrastive learning (GraphCL) is adopted to learn general and robust streamer and user representations. Apart from these learned self-supervised embeddings, we further extend behavior sequences through metapaths with the graph structural information and optimize the representations in an end-to-end manner with online recommendation model. Both of the self-supervised and end-to-end learning schemes help model jump out of the specific historical gifting behaviors for potential preferences exploration and largely enrich the behavior representation. Finally, to meet the low latency requirements of the online serving system, we propose a decoupled graph offline training and online inference strategy. MMBee has now been deployed on the live-streaming recommendation system of Kuaishou, serving millions of active users every day.

Overall, our contributions are shown as follows:
\begin{itemize}[leftmargin=*,topsep=1pt]
\vspace{-0.05cm}
\item The proposed Multi-modal Fusion with Learnable Query (MFQ) module leverages the dynamic multimodal content of live streaming and captures the distinct characteristics among streamers.
\item Graph-guided Interest Expansion (GIE) module largely enriches the observed history behaviors of users and streamers with both self-supervised graph representation learning and metapath-based behavior expansion to alleviate the sparsity problem.
\item We validate the effectiveness of MMBee through extensive offline experiments on Kuaishou’s 3 billion scale industrial dataset and public dataset. Online A/B tests further show that MMBee brings significant online benefits and we build efficient industrial infrastructure to deploy MMBee on the real-world online live streaming recommendation.
\end{itemize}

\vspace{-0.1cm}

\section{Related Work}
\vspace{-0.1cm}
\subsection{Live Streaming Gifting Recommendation}
\vspace{-0.1cm}

Existing works on live streaming gifting recommendation systems primarily view the whole live room as recommendation target and model the interaction between streamers and viewers only with categorical data. For instance, MARS \cite{lai2020live} proposes a novel recommendation scenario called Multi-Stream Party (MSP) and designs two-phase methods to jointly maximize the reciprocal response of donations and optimize MSP personal satisfaction. LSEC-GNN \cite{yu2021leveraging} models the live stream e-commerce scenario using GNN and fully leverages the interaction information among streamers, users, and products. However, previous research ignores that dramatic content changes can occur even within the same live room thus it is vital to make full use of the multi-modal feature in live streaming. Aiming to solve this issue, MTA \cite{xi2023multimodal} introduces a novel orthogonal projection model to capture the cross-modal information interaction of real-time content. However, MTA formulates the gifting prediction task as a time series prediction problem and neglects the personalization modeling of users' interests. In conclusion, there still exists great room for improvement in existing methods for live-streaming gifting prediction. 

\subsection{Personalized Recommendation}
\vspace{-0.1cm}
The most widely adopted personalized recommendation methods in the industry are based on deep neural networks. For instance, DIN \cite{zhou2018deep} models users' diverse interests in different target items by introducing attention mechanisms. SIM \cite{pi2020search} proposes an online two-stage retrieval method that models relevant behaviors from a user's long-term history based on the features of the current candidate item. However, in live-streaming gifting scenarios, it is challenging to achieve satisfactory results with these methods due to the sparsity issue in streamers and user interactions. Recently, several works combined with GNNs have introduced multimodal features to enrich the embedding of graph nodes. For instance, MMGCN \cite{wei2019mmgcn} captures user preferences across different modalities by constructing a modal-specific user-item bipartite graph. EgoGCN \cite{chen2022breaking} introduces a novel EGO fusion operation that enables inter-modal message spreading. However, the aforementioned methods all rely on recursive graph convolution to study the node embedding, which can result in exponential computation cost and significant inference latency, especially in live streaming gifting recommendation scenarios where the model needs to handle millions of nodes and ensure low latency during inference. Therefore, it is crucial to design an efficient graph architecture for training and inference.

\section{Preliminaries}
\vspace{-0.1cm}

In live streaming platforms, we use users to represent the viewers who watch live streaming and use authors to represent streamers. $V_u = \left \{ u_1,u_2,\cdots ,u_m \right \} $ is the set of users and $V_a = \left \{ a_1,a_2,\cdots ,a_k \right \} $ is the set of authors who are broadcasting at the current time, where $m$ is the numbers of users and $k$ is the numbers of authors. Previous studies treat the whole live streaming room as the recommendation target while ignoring the real-time change of streaming content. Thus, different from traditional recommendation tasks, we divide each live room into multiple consecutive 30s live segments and the live segment of author $a$ at the current time is denoted by $\delta_a$. We formulate that all live streaming segments of the current moment are the recommendation target and $M_{a}=\{v_a, s_a, t_a\}$ is the multi-modal raw data tuple, where $v_a,s_a$ and $t_a$ represent the visual frames, speech and comment text gathered from frame $\delta_a$. Given a set of triples $\left \langle  u_j, a_j, y_j\right \rangle $, $y_j = 1$ means that $u_j$ send gift to $a_j$, otherwise $y_j = 0$. Thus, the gift-through-rate (GTR) prediction problem is to predict whether user $u_i$ will send gift to $a_i$ given the multi-modal raw data $M_{a_i}=\{v_{a_i}, s_{a_i}, t_{a_i}\}$ in the current live streaming segment $\delta_{a_i}$: 
\begin{small}
\begin{equation}
p = f\left( a_i, u_i, M_{a_i} \right)
\label{gtrobjective}
\end{equation}
\end{small}
where $p$ is termed as the gift through rate (GTR) and $f(\cdot)$ is the GTR prediction model. In this work, we choose SIM \cite{pi2020search} as our foundational model, considering its widespread usage in the industry and its online efficiency and effectiveness. The objective function utilized in our method is the negative log-likelihood function, which is defined as follows: 

\begin{small}
\begin{equation}
L=-\frac{1}{N}  \sum_{i=1}^{N} (y_i \log p_i + (1-y_i) \log (1-p_i)))
\end{equation}
\end{small}
where $y_i$ denotes the ground truth label indicating whether current segment gets donation and $p_i \in \left [0,1 \right]$ is the predicted GTR.

\begin{figure*}[h]
\centering
\includegraphics[width=.99\textwidth]{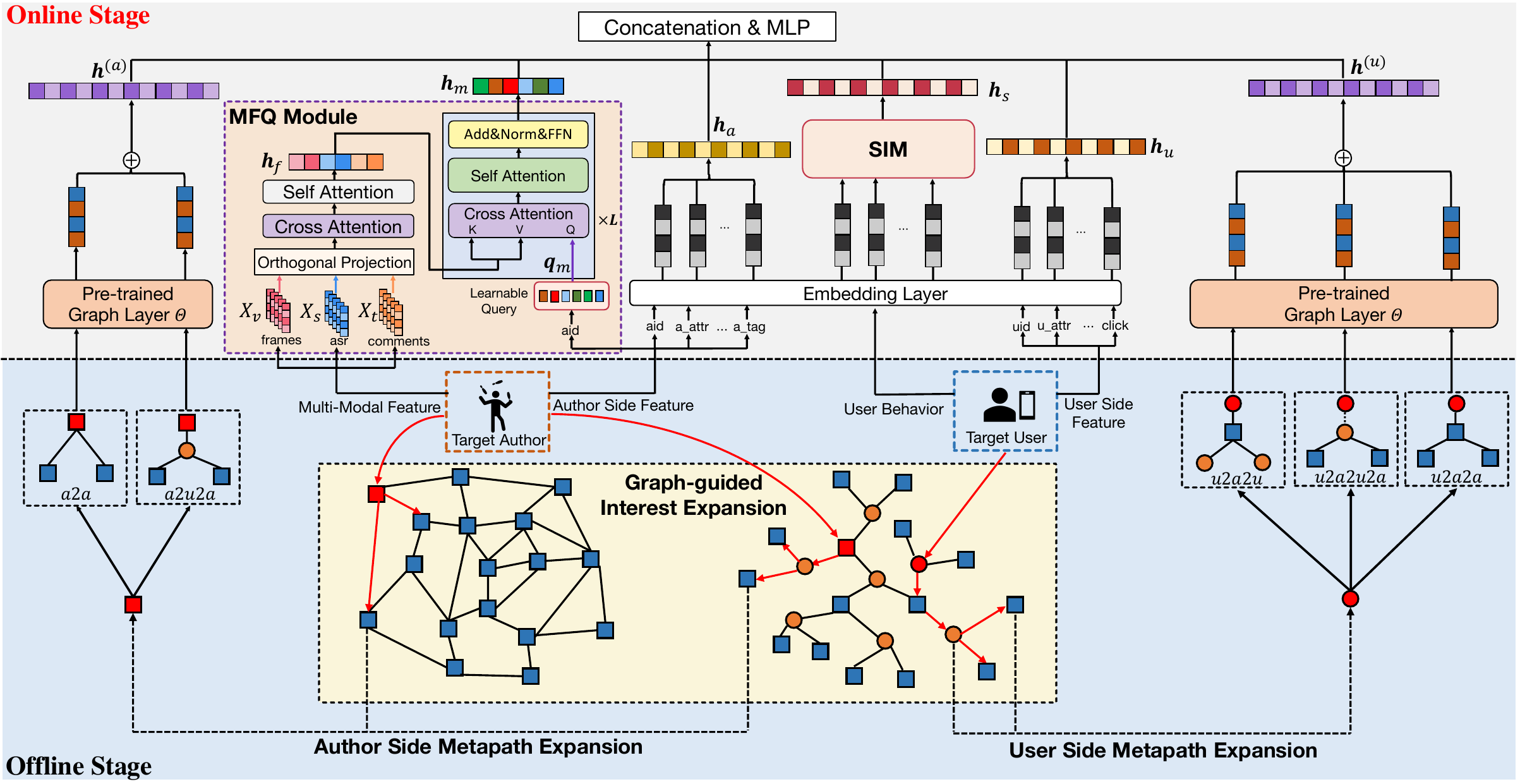}
\caption{\textbf{The overall framework of MMBee, consists of two stages: (i) the offline Graph-guided Interest Expansion (GIE) stage conducts the behavior expansion based on the target user and author; (ii) the online GTR prediction stage aggregates the real-time multi-modal content and expanded behavior for end-to-end training.}}
\label{fig4}
\vspace{-0.5cm}
\end{figure*}
 
\section{Multi-modal Fusion with Learnable Query}
For each live streaming segment, three frames are evenly sampled from each segment and necessary filtering process is conducted to clean the gathered ASR (Automatic Speech Recognition) and comment text. Then, we extract the multi-modal feature of raw data with Kuaishou's internal pre-trained 8 billion parameters multi-modal model K7-8B\footnote{\url{https://ir.kuaishou.com/news-releases/news-release-details/kuaishou-receives-leading-innvoation-digitial-economy-and-other}} and the extracted multi-modal feature sequences tuple of visual, speech and comment at the current moment of author $a$ are represented with $X_v$, $X_s$ and $X_t$, respectively.

Since processing and integrating information from different modalities is quite important \cite{yu2017multi, Kim2017, ben2019block}, we propose a multi-modal fusion with learnable queries to ensure efficient modality interactions. Inspired by \cite{xiao2022complementary, xi2023multimodal}, we adopt the orthogonal projection (OP) operation to maximize the complementation effects between different modalities. For example, take $X_v$ as target modality, we calculate the relevant scores between the visual modality $X_v$ with another two modalities by using correlation operations:
\begin{small}

\begin{equation}
\begin{aligned}
Corr_{vs} = \mathrm{Softmax}(X_v \cdot X_s)\\
Corr_{vt} = \mathrm{Softmax}(X_v \cdot X_t)
\end{aligned}
\end{equation}
\end{small}
where $\mathrm{Softmax}(\cdot)$ is the softmax operation. Then, the irrelevant parts are obtained through $1-x$ operation. Finally, the fused latent feature of visual modality $Y_v$ is performed with:
\begin{small}

\begin{equation}
\begin{aligned}
Y_v = OP(X_v, X_s, X_t) = X_v + X_s \cdot(1- Corr_{vs} ) \\
+ X_t \cdot(1- Corr_{vt} )
\end{aligned}
\end{equation}
\end{small}
Note that $1-Corr$ represents the dissimilarity vector that measures the difference between two modes’ representation. It helps to preserve the parts of other modalities that are orthogonal to the target modality and remove duplicate information to prevent redundancy.

Then, as shown in the online stage of Figure \ref{fig4}, we utilize the orthogonal latent features in a hybrid fusion \cite{song2021multimodal} manner applied with cross-attention and self-attention \cite{vaswani2017attention} alternately. 
The fused feature $\boldsymbol{h}_f$ is gotten with:
\begin{small}
\begin{equation}
\begin{gathered}
\boldsymbol{h}_v = \mathrm{CrossAttention}(X_v\boldsymbol{W}^Q_v, Y_v\boldsymbol{W}^K_v, Y_v\boldsymbol{W}^V_v), Y_v = OP(X_v, X_s, X_t)
\\
\boldsymbol{h}_s = \mathrm{CrossAttention}(X_s\boldsymbol{W}^Q_s, Y_s\boldsymbol{W}^K_s, Y_s\boldsymbol{W}^V_s), Y_s = OP(X_s, X_t, X_v)
\\
\boldsymbol{h}_t = \mathrm{CrossAttention}(X_t\boldsymbol{W}^Q_t, Y_t\boldsymbol{W}^K_t, Y_t\boldsymbol{W}^V_t), Y_t = OP(X_t, X_s, X_v) \\
\boldsymbol{h}^{'}_f = \boldsymbol{h}_v \oplus  \boldsymbol{h}_s \oplus  \boldsymbol{h}_t
\\
\boldsymbol{h}_f = \mathrm{SelfAttention}(\boldsymbol{h}^{'}_f\boldsymbol{W}^Q_f, \boldsymbol{h}^{'}_f\boldsymbol{W}^K_f, \boldsymbol{h}^{'}_f\boldsymbol{W}^V_f)
\end{gathered}
\end{equation}
\end{small}

However, the fused feature $\boldsymbol{h}_f$ can only reflect the content-level representation, thus lacking the connection to distinctive characteristics across various types of authors. To address this issue, we produce several learnable query\cite{zhu2020deformable, li2023blip} tokens $\boldsymbol{q}_m \in \mathbb{R}^{N \times d}$ to extract streamer-aware content patterns. Note that each author keeps a set number of learnable query embeddings which are randomly initialized. $N$ represents the number of query tokens for each author. The learnable query first interacts with fused multi-modal features through cross-attention layers as: 
\begin{small}
\begin{equation}
\begin{aligned}
\boldsymbol{h}^{'}_m = \mathrm{CrossAttention}(\boldsymbol{q}_m\boldsymbol{W}^Q_c, \boldsymbol{h}_f \boldsymbol{W}^K_c, \boldsymbol{h}_f\boldsymbol{W}^V_c) 
\end{aligned}
\end{equation}
\end{small}

Then the queries interact with each other through self-attention layers to fuse the necessary information among different patterns:

\begin{small}
\begin{equation}
\begin{aligned}
\boldsymbol{h}_m = \mathrm{SelfAttention}(\boldsymbol{h}^{'}_m\boldsymbol{W}^Q_s, \boldsymbol{h}^{'}_m\ \boldsymbol{W}^K_s, \boldsymbol{h}^{'}_m \boldsymbol{W}^V_s)
\end{aligned}
\end{equation}
\end{small}

The multi-modal fusion module benefits from the learnable queries in two major aspects: 1) Each author has learnable tokens that store their specific highlight content patterns. The tokens can be activated at certain moments of awesome content, which is quite useful for gifting prediction. 2) These queries help align the multimodal representations with the ID embedding based recommendation space, thereby maximizing their mutual information. Consequently, the integration of learnable queries further enhances model's ability to capture real-time content.

\section{Graph-guided Interest Expansion}

\subsection{User-to-Author and Author-to-Author Graph}

\begin{figure}[h]
\centering
\includegraphics[width=.45\textwidth]{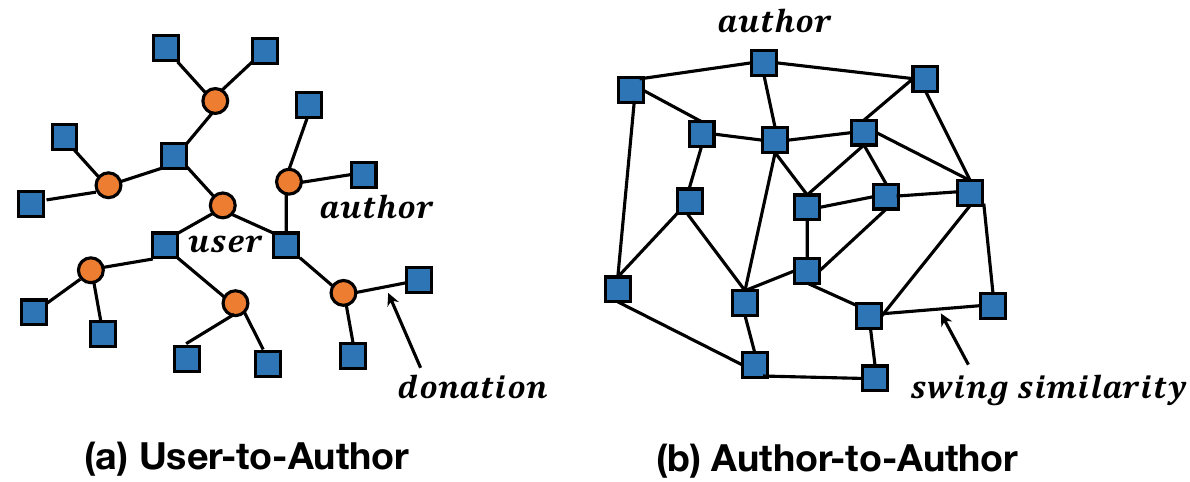}
\vspace{-0.4cm}
\caption{\textbf{User-to-author and author-to-author donation graph construction with donation history.}}
\label{fig2}
\vspace{-0.4cm}
\end{figure}

Based on the users' donation history, we first construct a User-to-Author(U2A) graph $G_1(V_u \cup V_a,E_1)$ that represents the correlation between users and authors, where $V_u$ and $V_a$ are the sets of users and authors respectively and $E_1$ represents the donation relationship between users and authors. As illustrated in Figure \ref{fig2} (a), the circle represents the user, and the square represents the author. If a user has previously made a donation, an edge exists between the user and the donated author in this graph. The weight of the edge is the amount of donated money and an author node has the attribute of aggregated multi-modal feature. In this way, the large User-to-Author graph is constructed. 

Based on the aforementioned U2A graph, we further construct the Author-to-Author (A2A) Graph $G_2(V_a,E_2)$ to represent the interdependence among authors, where $E_2$ denotes the Swing similarity \cite{yang2020large} relationship among authors. In this graph, each node represents an author, and the edge weight represents the Swing similarity between the authors. The similarity between author $i$ and author $j$ is given below:
\begin{small}
\begin{equation}
s(i, j)=\sum_{u \in U_i \cap U_j} \sum_{v \in U_i \cap U_j} \frac{1}{\alpha+\left|I_u \bigcap I_v\right|}
\label{swing}
\end{equation}
\end{small}
where $U_i$ is the set of users who have made donations on author $i$ and $I_u$ is the set of authors that donated by user $u$.

A2U graph is established through donation relationship between users and authors. The design of edge weights and sampling strategies helps enrich the representations of authors who have a rich history of being donated. However, there are some new or cold-start authors. Their limited donation history makes it difficult to benefit from the A2U graph. Fortunately, A2A graph is built from the swing similarity defined in Equation \ref{swing}, which finds substitutable authors based on the substructures of user-author donation bi-partitive graph. It is useful for linking cold-start author to warm-start author and encouraging the engagement of cold-start authors, so A2A graph is quite necessary.

After constructing U2A and A2A graphs, we first leverage the graph node representation learning approach to train graph embedding layer in Section \ref{graph_node_training}. Next, we propose metapath based behavior expansion process to enrich sparse behavior sequences in Section \ref{behaviour_expansion}. To provide a precise demonstration of the abovementioned methods, we first establish the following definition:

\begin{dfn}[Metapath\cite{fan2019metapath}]
Metapath is defined as a relation sequence to capture the specific structural relation between objects. In A2U and A2A graph, we define five metapaths: three metapaths $\rho _{u2a2u}, \rho _{u2a2u2a}, \rho _{u2a2a}$ begin from target user, for example $\rho _{u2a2a}=User\to Author\to Author$ metapath indicates that user make donation to authors in U2A graph, and these authors further retrieve similar authors in A2A graph, and we define two metapath $\rho _{a2a}$ and $\rho _{a2u2a}$ which begin form target author.
\end{dfn}

\begin{dfn}[Metapath-guided Neighbors\cite{fan2019metapath}]
Given a node $o$ and a metapath $\rho$ (start from $o$) in the graph, the metapath-guided neighbors are defined as the set of all visited nodes when the node $o$  walks along the given metapath. We denote the $i$-th step neighbors of object $o$ as $\mathcal{N}_\rho^{(i)}(o)$. For example, give the metapath $\rho _{u2a2u}=User\to Author\to User$, we can get metapath-guided neighbors as $\mathcal{N}_{\rho_{u2a2u}}^{(1)}\left(u_t\right)=\left\{a_1, a_2\right\}$, $\mathcal{N}_{\rho_{u2a2u}}^{(2)}\left(u_t\right)=\left\{u_1, u_2, u_3\right\}$.
\end{dfn}

\subsection{Node Representation Pre-training with GraphCL}
\label{graph_node_training}

\begin{spacing}{1.0}
Previous studies \cite{yu2014personalized, catherine2016personalized, palumbo2017entity2rec, palumbo2018knowledge} have shown that graph node embedding algorithms are beneficial for recommendation systems for tackling data sparsity problem because these methods are able to effectively capture the user-author relatedness from graph structures. To leverage the connectivity information of the whole graph, we apply the graph contrastive learning (GraphCL) framework to train the graph embedding layer. Aiming to cluster similar nodes together while pushing away dissimilar ones, we loop through all nodes in the whole graph $G_1$ and obtain positive sample set $V_p$ through the metapath-guided neighbor process and the negative nodes set $V_n$ are sampled randomly. The positive and negative nodes are utilized with the Cross-Entropy loss $\mathcal{L}_{CE}$ and InfoNCE \cite{oord2018representation} $\mathcal{L}_{NCE}$ loss for optimizing the parameters of the node embedding layers. Algorithm \ref{graphtraining} shows the core of our approach and the trained graph node embedding implies the connectivity information from the whole graph. The InfoNCE loss is defined with Equation \ref{clloss}.
\end{spacing}
\begin{footnotesize}
\begin{equation}
\begin{aligned}
\mathcal{L}_{NCE} = -\frac{1}{|V_{p}|} \sum_{v_i \in V_{p}} \log \left(\dfrac{\exp{\Theta(v_t)^T\Theta(v_i)}}{\exp{\Theta(v_t)^T \Theta(v_i)} + \sum\limits_{v_j \in V_{n}}\exp{\Theta(v_t)^T\Theta(v_j)}}\right)
\end{aligned}
\label{clloss}
\end{equation}
\end{footnotesize}

\begin{algorithm}[t]
  \SetAlgoLined
    
  Initialize $\mathcal{L} \leftarrow 0$\;
  
  Graph $G_1(V_u \cup V_a, E_1)$, graph node embedding layers parameter $\Theta \in  \mathbb{R} ^{\left | V_u \cup V_a \right |\times d  }$, walks epoch $\gamma $\;
  \For{$i= 0 $ to $\gamma $}{
    $\mathcal{O} = \mathrm{Shuffle}(V_u \cup V_a)$\;

    \For{ $ v_t \in \mathcal{O} $}{
      $V_{p} \leftarrow \{\}$, $V_{n} \leftarrow \{\}$\;
      \If{$v_t \in V_u $}{
        $V_{p} \leftarrow \mathcal{N}_{\rho_{u2a2u}}^{(2)}\left(v_t\right)$\;
        $V_{n} \leftarrow V_{n} \cup RandomSample(V_u)$\;
      }
      \If{$v_t \in V_a $}{
        $V_{p} \leftarrow \mathcal{N}_{\rho_{a2u2a}}^{(2)}\left(v_t\right)$\;
        $V_{n} \leftarrow V_{n} \cup RandomSample(V_a)$\;
      }
    }
    $\mathcal{L} \leftarrow \mathcal{L}_{CE} + \lambda \mathcal{L}_{NCE} $\;
    $\Theta \leftarrow \Theta - \alpha  \frac{\partial \mathcal{L} }{\partial \Theta }  $\;
  }
    
  \KwOut{Trained graph node embedding layers parameter $\Theta$}
  \caption{GraphCL}
  \label{graphtraining}
\end{algorithm}
\subsection{Metapath-guided Behavior Expansion through End-to-End Training}

\label{behaviour_expansion}
When analyzing the node number distribution of the constructed A2U graph, we observe that the average outdegree of user nodes is 0.32. It becomes difficult for widely used behavior-based models like SIM to study meaningful representations and explore potential gifting preferences. Furthermore, the graph embedding in Section \ref{graph_node_training} is trained in a self-supervised manner which is not directly optimized for the recommendation model. To address these challenges, we expand the behavior sequence of the target user and author using various pre-defined metapaths \cite{fan2019metapath}. Due to the computation cost, we perform up to 3-hop neighbors on both U2A and A2A Graph. We enumerate all possible metapaths and five metapaths with the highest scores are selected using commonly used feature importance filtering methods as follows:

\begin{itemize}[leftmargin=*,topsep=1pt]
\item \textit{$\mathcal{N}_{\rho_{u2a2u}}^{(2)}\left(u_t\right)$} begins with the target user $u_t$ and follow this metapath. The retrieved behavior sequence is a set of users who share the same authors as the target user. Therefore, this metapath gets similar users who share the similar interests of the target user.

\item \textit{$\mathcal{N}_{\rho_{u2a2u2a}}^{(3)}\left(u_t\right)$} helps identify potential authors that may reflect the interest of the target user, excluding the authors they have already donated to in the past.

\item \textit{$\mathcal{N}_{\rho_{u2a2a}}^{(2)}\left(u_t\right)$} is based on the target user's donated authors history and it retrieves similar authors in the A2A graph to find similar authors with respect to the target user.

\item \textit{$\mathcal{N}_{\rho_{a2a}}^{(1)}\left(a_t\right)$} begins with the target author $a_t$, it retrieves the similar authors in the A2A graph. Therefore, this metapath helps obtain similar authors to the target author. 

\item \textit{$\mathcal{N}_{\rho_{a2u2a}}^{(2)}\left(a_t\right)$} indicates that a group of users donates to the target author in the U2A graph, and these users subsequently donate to another group of authors. Therefore, this metapath helps identify potential interest authors for the target author.
\end{itemize}

Based on these metapath-guided neighbors, we significantly enrich the behavior sequence of the target user and author. During the offline GIE stage, we store the pre-aggregated embeddings of the metapath-guided expanded neighbors of each user and author on the graph into memories or key-value databases to be further utilized in the online training stage.

In order to eliminate the gap between pre-trained node representation and online recommendation model, we gather the expanded sequences and optimize them with GTR prediction objective in recommendation model for end-to-end training. The generation of user side expanded graph representation $\mathbb{E}^{(u)} $ can be formalized as:
\begin{small}
\begin{equation}
\mathbb{E}^{(u)} = \bigl\{\Theta(v_i)| v_i \in \mathcal{N}_{\rho_{u2a2u}}^{(2)}\left(u_t\right) \cup \mathcal{N}_{\rho_{u2a2u2a}}^{(3)}\left(u_t\right) \cup \mathcal{N}_{\rho_{u2a2a}}^{(2)}\left(u_t\right) \bigl\}
\end{equation}
\end{small}
And the generation of author side expanded graph representation $\mathbb{E}^{(a)} $ can be formalized as:
\begin{small}
\begin{equation}
\mathbb{E}^{(a)} = \bigl\{\Theta(v_i)| v_i \in \mathcal{N}_{\rho_{a2a}}^{(1)}\left(a_t\right) \cup \mathcal{N}_{\rho_{a2u2a}}^{(2)}\left(a_t\right) \bigl\}
\end{equation}
\end{small}
where $\Theta(\cdot)$ represents the graph node embedding layer operation and author multi-modal attribute retrieval operation. Then we implement the mean pooling and concatenate operation in the recommendation model to get the final graph embedding $\boldsymbol{h}^{(u)}$ and $\boldsymbol{h}^{(a)}$ for end-to-end training:
\begin{small}

\begin{equation}
\boldsymbol{h}^{(u)} = \mathrm{MeanPooling}(\mathbb{E}^{(u)}) \quad
\boldsymbol{h}^{(a)} = \mathrm{MeanPooling}(\mathbb{E}^{(a)})
\end{equation}
\end{small}

\subsection{System Deployment}
\label{System_Deployment}
\begin{figure}[h]
\vspace{-0.4cm}
\centering
\includegraphics[width=.49\textwidth]{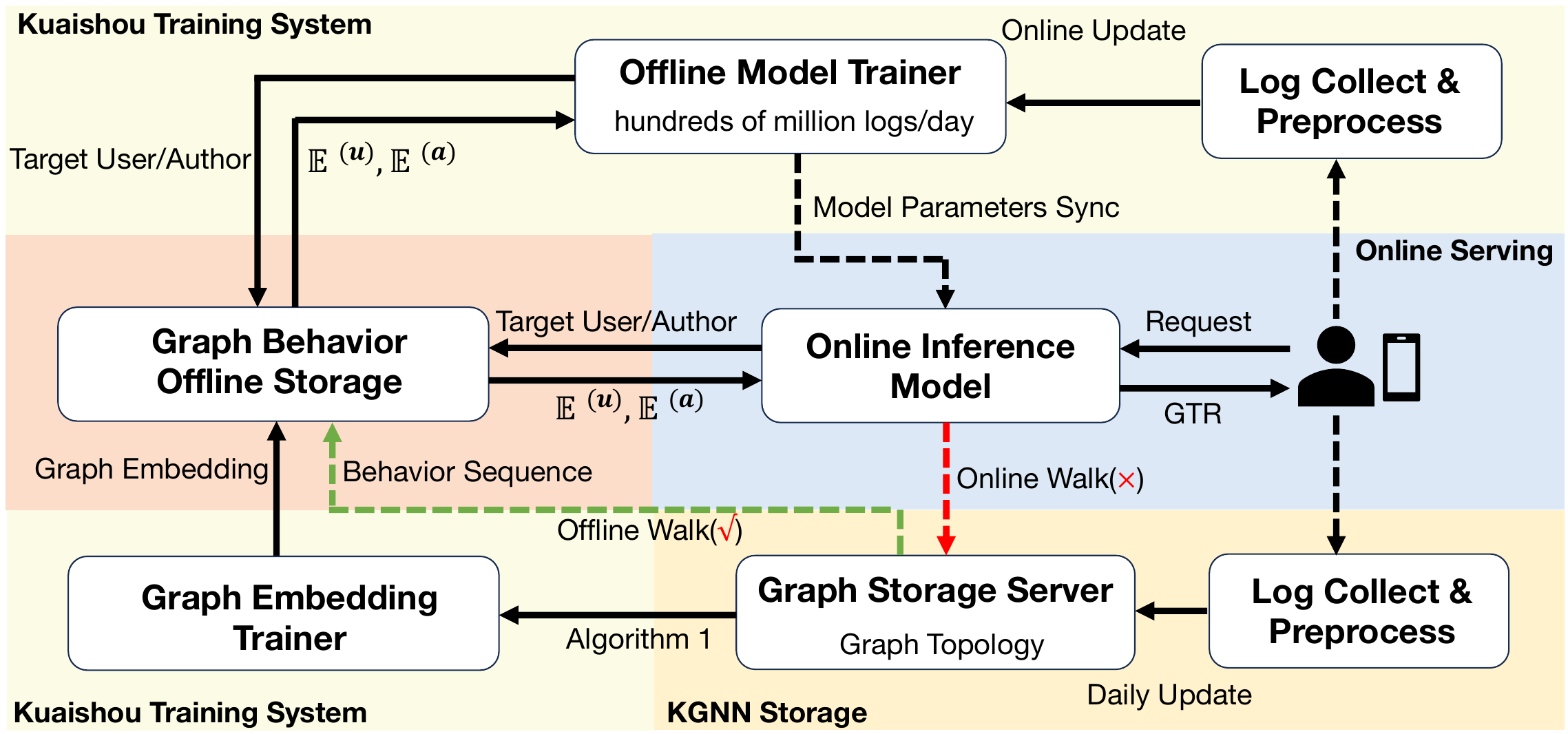}
\caption{\textbf{The deployment of MMBee in online live streaming GTR prediction system.}}
\label{fig6}
\vspace{-0.5cm}
\end{figure}

As shown in Figure \ref{fig6}, our recommendation model and graph embedding layer are trained on Kuaishou's large-scale distributed training system. Each day, hundreds of millions of users visit Kuaishou, actively watching and interacting with live-streaming content, resulting in the generation of hundreds of millions of logs for watching and interaction. These logs are collected, preprocessed in real-time, and utilized for training the model. Our training system incrementally updates the model parameters by incorporating the latest user-author interactions, multi-modal content features, and trained graph embedding. The trained parameters are synchronized to the online inference model for online serving. To train graph embedding, we first gather the users' historical donation behavior and utilize it to build the User-Author and Author-Author donation graphs. The topology of these two graphs is stored in a key-value based storage system called KGNN\footnote{\url{https://www.jiqizhixin.com/articles/2020-12-08}}. Then the graph embedding trainer requests the KGNN server with Algorithm \ref{graphtraining} for training the node embedding layer and the KGNN storage updates once a day. 

During the training and inference processes of the recommendation model, it needs to request the metapath-guided neighbors of the target user and author. As shown by the red dashed line in Figure \ref{fig6}, one approach is to dynamically request the KGNN storage. However, this method can impose significant computational overhead on the KGNN server and result in great time delays when walking on the entire graph. To address this issue, as shown by the green dashed line in Figure \ref{fig6}, we apply the pre-requested expansion manner and store the metapath-guided neighbors of all nodes in the graph in the Graph Behavior Offline Storage in advance. As a result, the online recommendation model can directly access the Graph Behavior Offline Storage to retrieve the sequence without having to walk on the graph.

\section{Experiment}
\subsection{Dataset}

\subsubsection{Kuaishou Dataset} We first test our method on company internal dataset called Kuaishou Dataset. It includes about 3 billion user interaction logs with live-streaming content in Kuaishou App. This dataset is collected as follows: We first apply a 30s sliding window to generate the streaming segment samples. If the user requests the recommendation service and makes a donation at time $t$, then only the segment containing $t$ will be taken as the positive training sample while other samples will be ignored. On the contrary, if the recommended live broadcast has impressed but users' donation behavior does not occur until exiting, the segment when user exit will be adopted as negative sample \cite{liang2024ensure}. With this process, the sparsity of of Kuiashou dataset is 99.969\% which is reasonable. Kuaishou dataset is composed of two parts: $D_{train}$ and $D_{test}$, where $D_{train}$ is users' real interaction logs from 7 days of all live streaming content during that period for the training phase. The $D_{test}$ is sampled from the following one-day logs after $D_{train}$ is collected, which is used to test model's performance.

\subsubsection{Public Dataset} To prove the effectiveness of our proposed MFQ and GIE module, we also compare our method on two public short video recommendation datasets: TikTok and MovieLens. The statistics of datasets are shown in Table \ref{publicdataset}.
\begin{table}[H]
\renewcommand\thetable{1}
    \centering
    \vspace{-8pt}
    \caption{
    The statistics of public datasets. V, A, and T represent the dimensions of visual, acoustic and textual features.
    }
    \vspace{-6pt}
    \label{publicdataset}
    \setlength{\tabcolsep}{3.1pt}
    \scalebox{0.9}{
    \begin{tabular}{p{1.3cm}<{\raggedright}|p{1.6cm}<{\centering}|p{0.8cm}<{\centering}|p{0.8cm}<{\centering}|p{1cm}<{\centering}|p{0.6cm}<{\centering}|p{0.6cm}<{\centering}|p{0.6cm}<{\centering}}
    \hline 
    Dataset & \#Interactions & \#Items &  \#Users & Sparsity & V & A & T \\ 
    \hline
    Tiktok & 726,065 & 76,085 &  36,656 & 99.97\% & 128 & 128 & 128 \\
    Movielens & 1,239,508 & 5,986 &  55,485 & 99.63\% & 2048 & 128 & 100 \\
    \hline
    \end{tabular}}
    \vspace{-10pt}
\end{table}

\textbf{Movielens\footnote{\url{https://grouplens.org/datasets/movielens/}}} \cite{harper2015movielens} is a widely used dataset \cite{sun2019bert4rec, kang2018self, sankar2020dysat, wu2022graph} for the recommendation task. The raw data is initially acquired by collecting movie descriptions from Movielens-10M and crawling the corresponding trailers from YouTube. Textual features are subsequently extracted from the descriptions using the Sentence2Vector \cite{arora2017simple}. For visual modality, key frames are initially extracted from the retrieved videos and then processed by a pre-trained ResNet50 model [9] to obtain visual features. The acoustic features are obtained using VGGish [12], following a soundtrack separation procedure implemented with the FFmpeg software.

\textbf{TikTok\footnote{\url{http://ai-lab-challenge.bytedance.com/tce/vc/}}} is published by TikTok, a micro-video sharing platform that enables users to create and share micro-videos with durations ranging from 3 to 15 seconds. TikTok comprises users, micro-videos, and their interactions, such as clicks. The features of the micro-videos in each modality are extracted and made available without providing the raw data. Specifically, the textual characteristics are extracted from the micro-video captions provided by users.

\subsection{Baseline}
On the Kuaishou dataset, we choose two widely used baselines MMoE \cite{ma2018modeling} and SIM \cite{pi2020search} for comparison. We evaluate the performance of our method by comparing it with the following recommendation method that is integrated with MMoE and SIM:
\begin{table*}[htbp]
\renewcommand\thetable{2}
    \centering
    \caption{
    Performances of different methods on Kuaishou dataset. $^*$ represents the absolute improvement.
    }
    \label{overall}
    \vspace{-8pt}
    \setlength{\tabcolsep}{3.1pt}
    \scalebox{0.95}{ 
    \begin{tabular}{p{3.5cm}<{\raggedright}|p{1.5cm}<{\centering}p{1.7cm}<{\centering}p{1.5cm}<{\centering}p{1.7cm}<{\centering}p{1.5cm}<{\centering}p{1.7cm}<{\centering}}
    \hline 
     \multicolumn{1}{c|}{\multirow{2}{*}{\textbf{Methods}}} & \multicolumn{6}{c}{\textbf{GTR}  } \\
    \multicolumn{1}{c|}{}             &   \textbf{AUC} & \textbf{Impr.}$^*$ & UAUC & \textbf{Impr.}$^*$  & GAUC  & \textbf{Impr.}$^*$    \\ 
    \hline
     \textbf{MMoE} \cite{ma2018modeling} &0.956230 & -& 0.730186 & - & 0.746711 & -\\ 
     \textbf{MMoE+BDR} \cite{zhang2021deep} & 0.956908 & +0.0678 \% &   0.730625 &+0.0439 \% &  0.747136  & +0.0425 \% \\ 
    \textbf{MMoE+MTA} \cite{xi2023multimodal}&  0.957095  &+0.0865 \% & 0.731450 & +0.1264 \%&  0.747327 & +0.0616 \%\\ 
     \textbf{MMoE+EgoFusion} \cite{chen2022breaking} & 0.956952   & +0.0722 \%  &   0.731418  &  +0.1232 \% &  0.747275  & +0.0564 \% \\ 
      \cline{1-7} %
     \textbf{MMoE+MFQ} & 0.956902   & +0.0672 \%  &   0.731975  &  +0.1789 \% &  0.747275  & +0.1764 \% \\ 
     \textbf{MMoE+GIE} & 0.957064   & +0.0834 \%  &   0.733853  &  +0.3667 \% &  0.751239  & +0.4528 \% \\ 
    \textbf{MMoE+Ours(MFQ+GIE)} & \textbf{0.95723}   & \textbf{+0.1001 \%} &   \textbf{0.735776} &  \textbf{+0.5590 \%} & \textbf{0.753017} & \textbf{+0.6306 \%}\\ 
    \hline
    \hline
    \textbf{SIM} \cite{pi2020search} & 0.958656 & - &   0.732239 & - &  0.748383  & - \\ 
    \textbf{SIM+BDR} \cite{zhang2021deep} & 0.958419 & -0.0237 \% &   0.734757 &+0.2518 \% &  0.750684  & +0.2301 \% \\ 
    \textbf{SIM+MTA} \cite{xi2023multimodal} &  0.958867  &+0.0211 \% & 0.734921 & +0.2682 \%&  0.750802 & +0.2419 \%\\ 
     \textbf{SIM+EgoFusion} \cite{chen2022breaking} & 0.959387   & +0.0085 \%&   0.735608 &  +0.3369 \% &  0.751669  & +0.3286 \%\\ 
      \cline{1-7} %
     \textbf{SIM+MFQ} & 0.959202   & +0.0546 \%  &   0.735717  &  +0.3478 \% &  0.751780  & +0.3397 \% \\ 
     \textbf{SIM+GIE} & 0.959802   &  +0.1146 \%  &   0.738309  &  +0.6070 \% &  0.755154  & +0.6771 \% \\ 
     \textbf{SIM+Ours(MFQ+GIE)} & \textbf{0.960302}   & \textbf{+0.1646 \%}  &   \textbf{0.743678} &  \textbf{+1.1439 \%} & \textbf{0.76044} & \textbf{+1.2057 \%} \\ 

    \textit{p-value} & \multicolumn{2}{c}{$1.02e^{-3}$} &    \multicolumn{2}{c}{$2.01e^{-3}$} &  \multicolumn{2}{c}{$5.12e^{-3}$} \\ 
    \hline
    \end{tabular}}
    \vspace{-4pt}
\end{table*}

\begin{table*}[htbp]
\renewcommand\thetable{3}
    \centering
    \vspace{-2pt}
    \caption{
    Performances of different methods on Tiktok and Movielens datasets. 
    }
    \label{public}
    \vspace{-5pt}
    \setlength{\tabcolsep}{3.1pt}
    \scalebox{0.95}{
    \begin{tabular}{p{2.2cm}<{\raggedright}|p{1.8cm}<{\centering}p{1.8cm}<{\centering}p{1.8cm}<{\centering}|p{1.8cm}<{\centering}p{1.8cm}<{\centering}p{1.8cm}<{\centering}}
    \hline 
     \multicolumn{1}{c|}{\multirow{2}{*}{\textbf{Methods}}} & \multicolumn{3}{c|}{\textbf{TikTok}} & \multicolumn{3}{c}{\textbf{Movielens}} \\
    \multicolumn{1}{c|}{}             &   Recall@10 & Precision@10& NDCG@10 & Recall@10& Precision@10  &  NDCG@10   \\ 
    \hline
    \textbf{NGCF} \cite{wang2019neural} & 0.0292 &0.0045 &0.0156 &0.1198 &0.0289 &0.0750  \\ 
    \textbf{LightGCN} \cite{he2020lightgcn} & 0.0448 &0.0082 &0.0261 &0.1992 &0.0479 &0.1324 \\ 
    \textbf{MMGCN} \cite{wei2019mmgcn} &  0.0544  & 0.0089 & 0.0297 & 0.2028&  0.0506 & 0.1361\\ 
    \textbf{GRCN} \cite{wei2020graph} & 0.0392   & 0.0065 &  0.0221  &  0.1402 &  0.0338  & 0.0882\\ 
    \textbf{EgoGCN} \cite{chen2022breaking} & \underline{0.0569}   & \underline{0.0093} &  \underline{0.0330} &  0.2155 & \underline{0.0524} & \underline{0.1444}\\ \hline  
    \textbf{DIN} \cite{zhou2018deep} & 0.0403   & 0.0074 &  0.0235 &  0.1372 & 0.0330 & 0.0912\\
    \textbf{SASRec} \cite{kang2018self} & 0.0435   & 0.0043 &  0.0215 &  0.1914 & 0.0191 & 0.1006\\
    \textbf{SIM} \cite{pi2020search} & 0.0413   & 0.0079 &  0.0245 &  0.1470 & 0.0429 & 0.1011\\
    \textbf{MMMLP} \cite{liang2023mmmlp} & 0.0509   & 0.0081 &  0.0297 &  0.1842 & 0.0484 & 0.1328\\
    \textbf{MMSSL} \cite{pi2020search} & 0.0553   & 0.0055 &  0.0299 &  \textbf{0.2482} & 0.0170 & 0.1113\\ \hline
    \textbf{Ours} & \textbf{0.0605} & \textbf{0.0097} &  \textbf{0.0347} & \underline{0.2317} &  \textbf{0.0566}  & \textbf{0.1573} \\ 
    \textit{p-value} & $1.29e^{-5}$ &  $6.23e^{-6}$ &   $7.29e^{-5}$ &  $2.75e^{-5}$ &   $2.81e^{-3}$  &  $1.61e^{-2}$ \\ 
    \hline
    \end{tabular}}
    \vspace{-6pt}
\end{table*}

\begin{itemize}[leftmargin=*,topsep=1pt]
\item \textbf{BDR} \cite{zhang2021deep} consists of User-to-User and Author-to-Author graphs, enabling simultaneous prediction from both perspectives. 

\item \textbf{MTA} \cite{xi2023multimodal} leverages multimodal time-series analysis to effectively integrate information from different modalities. This approach does not consider modeling personalized preferences.

\item \textbf{EgoFusion} \cite{chen2022breaking} allows the spread of inter-modal messages in EgoGCN. In our work, we apply the Ego fusion operation to the multi-modal feature of node attribution to generate the multi-modal embedding and we exclude the MFQ module for a fair comparison in this baseline.

\end{itemize}

On the public dataset, we compare the performance of our method with the following GCN-based models:

\begin{itemize}[leftmargin=*,topsep=1pt]
\item \textbf{NGCF} \cite{wang2019neural} exploits high-order connectivity and collaborative signal by propagating embeddings on user-item graph structure.

\item \textbf{LightGCN} \cite{he2020lightgcn} remove feature transformation and nonlinear activation from standard GCNs to construct a lightweight structure for collaborative filtering.

\item \textbf{MMGCN} \cite{wei2019mmgcn} captures modality-specific user preferences and integrates them to form user representations, which is used to evaluate their affinities towards the content features of the items.

\item \textbf{GRCN} \cite{wei2020graph} refines the user-item bipartite sub-graphs for different modalities and adjusts the representation of the user and item accordingly to improve the prediction of their interactions.

\item \textbf{EgoGCN} \cite{chen2022breaking} improves the user-item interactions through an effective graph fusion approach called EGO fusion.
\end{itemize}

We also further compare the performance of our method with well-known recommendation methods besides graph and recent methods integrated with multi-modal features:
\begin{itemize}[leftmargin=*,topsep=1pt]

\item \textbf{DIN} \cite{zhou2018deep} captures temporal interests from history behavior sequence with GRU and attentional update gate.

\item \textbf{SASRec} \cite{kang2018self} is a classic transformer-based sequential recommender.

\item \textbf{SIM} \cite{pi2020search} models life-long behavior in the two cascading stages with General Search Unit (GSU) and Exact Search Unit (ESU).

\item \textbf{MMMLP} \cite{liang2023mmmlp} adapts MLP-Mixer for modelling multi-modal feature in  sequential recommendation .

\item \textbf{MMSSL} \cite{pi2020search} addresses the sparsity issue by introducing self-supervised tasks that maximize the mutual information between multiple content-augmented views.
\end{itemize}

\subsection{Evaluation Metrics}

For offline evaluation on Kuiashou dataset, we use the training set $D_{train}$ to train all methods and evaluate the performance of all methods on the test set $D_{test}$. We report the average performance over hours. We adopt three widely adopted metrics: AUC, UAUC and GAUC \cite{zhang2022keep} to evaluate the performance of different methods. AUC represents the probability that the score of a positive sample is higher than that of a negative sample, reflecting the ranking capability of a model. UAUC is the average of AUC values calculated for different users and GAUC is the weighted average of UAUC considering the impressions. They are defined as follows:
\begin{small}
\begin{equation}
\mathrm{UAUC}=\frac{1}{N}\sum_{i=1}^{N} \mathrm{AUC}_i \quad \mathrm{GAUC}=\dfrac{\sum_{i=1}^{N} impression_i \ast \mathrm{AUC}_i }{\sum_{i=1}^{N} impression_i}
\end{equation}
\end{small}

where $N$ refers to the number of active users in the testing set. UAUC and GAUC alleviate the bias among users and consider the effect of impression to evaluate the model's performance in a finer and fair manner.

\subsection{Overall Performance}

Table \ref{overall} shows the performance of all models on the Kuaishou dataset. Note that given the large number of users and samples in Kuaishou dataset, an improvement of 0.5\% in AUC, UAUC, and GAUC during offline evaluation holds significant value to bring obvious online gains for business. Table \ref{public} presents the performance of several competitors on public Tiktok and Movielens datasets.

\textbf{First, our method surpasses all baselines by a significant margin on Kuaishou dataset.} Our method MFQ significantly outperforms traditional live streaming recommendation models BDR and MTA in UAUC and GAUC for two main reasons. Firstly, BDR ignores the modeling of multi-modal content, while MTA lacks the connection to distinctive characteristics across various types of authors. In contrast, our MFQ successfully leverages the multi-modal content of the target live-streaming room and adopts learnable queries to extract streamer-aware content patterns. Additionally, our method GIE also outperforms the graph-based method EgoFusion which provides evidence that the metapath-guided behavior expansion process greatly enhances behavior representation and explores potential donation preferences.

\textbf{Secondly, our method exhibits generalizability to a common behavior-based model.} Our method has seamlessly integrated into two widely used behavior-based methods, MMoE and SIM, both of which demonstrate significant performance improvements. Moreover, MMBee is not limited to these two behavior-based models and can be easily adapted to other methods such as DIN \cite{zhou2018deep} and DIEN \cite{zhou2019deep} as well.

\textbf{Thirdly, our method is not restricted to gifting prediction tasks and it also proves effectiveness in multi-modal recommendation tasks.} As shown in Table \ref{public}, our method exhibits great improvement when compared to several strong multi-modal recommendation baselines. This gain mainly comes from two folds: (1) The metapath-guided neighbors in our method enable better capture of user preferences, but other graph-based methods only rely on implicit learning from graph embeddings. (2) The MFQ module enhances the fusion of multi-modal features from short videos and clusters different videos with learnable queries initialized with item embedding, thereby benefiting further performance improvement of the recommendation model. 
\subsection{Ablation Study}
\begin{table*}[htbp]
    \renewcommand\thetable{7}
    \centering
    \vspace{-2pt}
    \caption{
    Ablation Study on Graph and Mutli-modal level. The number in bold indicates a significant performance degradation.
    }
    \vspace{-6pt}
    \label{abliation}
    \setlength{\tabcolsep}{3.1pt}
    \scalebox{0.95}{
    \begin{tabular}{p{1.8cm}<{\centering}|p{1.8cm}<{\centering}|p{1.8cm}<{\centering}p{1.8cm}<{\centering}|p{1.8cm}<{\centering}p{1.8cm}<{\centering}|p{1.8cm}<{\centering}p{1.8cm}<{\centering}}
    \hline 
    \textbf{Category} & \textbf{Operator} & \textbf{AUC} &  \textbf{Impr.} & \textbf{UAUC}  & \textbf{Impr.} & \textbf{GAUC} & \textbf{Impr.} \\ 
    \hline
    - & SIM & 0.958656 & -0.1646\%  & 0.732239  & -1.1439\% & 0.748383 & -1.2057 \% \\
    \hline
    \multicolumn{1}{c|}{\multirow{7}{*}{\textbf{Graph}}} & $\boldsymbol{h_{u2a2u}(-)}$ & 0.959842 &  -0.0460 \% & 0.743492 & -0.0186 \% & 0.76014 & -0.0300 \% \\
    \multicolumn{1}{c|}{}  & $\boldsymbol{h_{u2a2u2a}(-)}$ & 0.959706 &  \textbf{-0.0596} \% & 0.738322 & \textbf{-0.5356 \%} & 0.755081 & \textbf{-0.5359 \%} \\
    \multicolumn{1}{c|}{}  & $\boldsymbol{h_{u2a2a}(-)}$ & 0.960162 &  -0.0140 \% & 0.743248 & -0.0430 \% & 0.75976 & -0.0680 \% \\
    \multicolumn{1}{c|}{}  & $\boldsymbol{h_{a2a}(-)}$ & 0.960002 &  -0.0300 \% & 0.742931 & -0.0747 \% & 0.759818 & -0.0622 \% \\
    \multicolumn{1}{c|}{}  & $\boldsymbol{h_{a2u2a}(-)}$ & 0.959462 &  \textbf{-0.0840 \%} & 0.738378 & \textbf{-0.5300 \%} & 0.754722 & \textbf{-0.5718 \%} \\
    \multicolumn{1}{c|}{}  & $\boldsymbol{\Theta(-)}$ & 0.959782 &  \textbf{-0.0520\%} & 0.736832 & \textbf{-0.6846 \%} & 0.752625 &\textbf{-0.7815 \%} \\
    \multicolumn{1}{c|}{}  & $\boldsymbol{h_{g}(-)}$ & 0.959202 &  \textbf{-0.1100\%} & 0.735608 & \textbf{-0.8070 \%} & 0.751669 &\textbf{-0.8771 \%} \\
    \hline
    \multicolumn{1}{c|}{\multirow{2}{*}{\textbf{Multi-modal}}} & $\boldsymbol{h_{m}(-)}$ & 0.959802 &  \textbf{-0.0500 \%} & 0.738309 & \textbf{-0.5369 \%} & 0.755154 & \textbf{-0.5286 \%} \\
    \multicolumn{1}{c|}{}  & $\boldsymbol{q_{m}(-)}$ & 0.960091 &  -0.0211\% & 0.740996 & -0.2682 \% & 0.758021 &-0.2419 \% \\
    \hline
    - & \textbf{Ours} & 0.960302 &  0.0000 \% & 0.743678 & 0.0000 \% & 0.76044 & 0.0000 \% \\
    \hline
    \end{tabular}}
    \vspace{-1pt}
\end{table*}

\textbf{Graph-level Ablation:} In order to investigate the importance of different metapath neighbors and the effect of graph embedding training, we remove five expanded sequences in turn and evaluate the performance of ablated graph embedding features. The results are presented in Table \ref{abliation}, where we use $\boldsymbol{(-)}$ to represent the removed part or feature. For example, $\boldsymbol{h_{u2a2u}(-)}$ means removing the metapath neighbors $\mathcal{N}_{\rho_{u2a2u}}^{(2)}\left(u_t\right)$ in recommendation model, $\boldsymbol{\Theta(-)}$ denotes removing the learned graph node embedding layers but remaining the expanded sequence and $\boldsymbol{h_{g}(-)}$ represents removing all features of graph modeling. From table \ref{abliation}, we can observe that $\boldsymbol{h_{g}(-)}$ drops -0.1100\% of AUC  and  $\boldsymbol{\Theta(-)}$ also leads to a significant drop in performance which means that the GIE modeling is a very important supplement to the observed history behaviors. This suggests that the explicit metapath-based behavior expansion process and implicit graph node embedding learning are all beneficial to model's performance. Furthermore, among five expanded behavior sequences, we observed the metapath of $\rho_{a2u2a}$ and $\rho_{u2a2u2a}$ are the most important sequences among them.

\noindent \textbf{Multi-modal Ablation:} We also investigate the influence of the multi-modal feature in MFQ module. Specifically, $\boldsymbol{h_{m}(-)}$ denotes removing all multi-modal content and $\boldsymbol{q_{m}(-)}$ represents removing the learnable query and cross attention. Table \ref{abliation} shows that when removing the multi-modal feature MMBee suffers significant performance drops. We further study the influence of different modalities and report the ablation results in Table \ref{rtable5}. We find visual modality has the most important impact, causing the most performance degradation when removed. The speech and comment modality have a lesser impact factor but still show an innegligible effect on the model’s overall performance.
\begin{table}[H]
\renewcommand\thetable{4}
    \centering
    \vspace{-8pt}
    \caption{
    Ablation study on different modality impact.
    \vspace{-4pt}
    }
    \label{rtable5}
    \setlength{\tabcolsep}{3.1pt}
    \scalebox{0.85}{
    \begin{tabular}{p{1.2cm}<{\raggedright}|p{0.5cm}<{\centering}|p{0.5cm}<{\centering}|p{0.5cm}<{\centering}|p{1.7cm}<{\centering}|p{1.8cm}<{\centering}|p{1.8cm}<{\centering}}
    \hline 
    \textbf{Methods} & $X_v$ & $X_s$	 &  $X_t$ & \textbf{AUC Impr.} & \textbf{UAUC Impr.}  & \textbf{GAUC Impr.}  \\ 
    \hline
    \textbf{MMBee} & $\surd$ & $\surd$ & $\surd$ & 0.0000\% & 0.0000\% & 0.0000\%  \\
    $\boldsymbol{X_{v}(-)}$ & - & $\surd$ & $\surd$ & -0.1101\% & -0.2069\% & -0.2939\%  \\
    $\boldsymbol{X_{s}(-)}$ & $\surd$ & - & $\surd$ & -0.1090\% & -0.1565\% & -0.1383\%  \\
    $\boldsymbol{X_{t}(-)}$ & $\surd$ & $\surd$ & - & -0.0839\% & -0.0933\% & -0.1790\%  \\
    \hline
    \end{tabular}}
    \vspace{-6pt}
\end{table}

\noindent \textbf{Hyperparameters Ablation:} We provide further experiment results about hyperparameters as follows:
\begin{itemize}[leftmargin=*,topsep=1pt]
\item \textbf{Dimension of MFQ.} We compare 32/64/128 dimensions of MFQ on Kuaishou dataset and the speed is tested on 20*Tesla T4 GPUs measured in examples/second. Table \ref{rtable4} shows that 64 dimension holds the best trade-off with computation efficiency and accuracy. 

\begin{table}[H]
\renewcommand\thetable{5}
    \centering
    \vspace{-8pt}
    \caption{
    The influence of dimension of MFQ.
    }
    \vspace{-4pt}
    \label{rtable4}
    \setlength{\tabcolsep}{3.1pt}
    \scalebox{0.85}{
    \begin{tabular}{p{2cm}<{\centering}|p{2.5cm}<{\centering}|p{2.cm}<{\centering}|p{2cm}<{\centering}}
    \hline 
    \textbf{Dimension} & \textbf{Speed} & \textbf{FLOPs}	 &  \textbf{AUC Impr.}\\ 
    \hline
    32 & 144.17K & 154.61M & 0.0000\% \\
    64 & 141.76K & 190.27M & 0.1744\% \\
    128 & 132.73K & 229.30M & 0.2105\% \\
    \hline
    \end{tabular}}
    \vspace{-10pt}
\end{table}

\item \textbf{Segment Length.} We additionally choose 10/20 consecutive live segments and compared them with 5 segments on Kuaishou dataset. Table \ref{rtable2} shows that 10 live segments get obvious gain but when it comes to 20 the further gain is modest. However, 10 segments significantly increase resource costs (including storage, training and serving) making it infeasible to deploy in production. So we use 5 segments in MMBee.

\begin{table}[H]
\renewcommand\thetable{6}
    \centering
    \vspace{-6pt}
    \caption{
    The influence of segments length.
    }
    \vspace{-4pt}
    \label{rtable2}
    \setlength{\tabcolsep}{3.1pt}
    \scalebox{0.85}{
    \begin{tabular}{p{1.cm}<{\centering}|p{1.6cm}<{\centering}|p{2.0cm}<{\centering}|p{1.6cm}<{\centering}|p{1cm}<{\centering}|p{1.cm}<{\centering}}
    \hline 
    \textbf{Length} & \textbf{AUC Impr.} & \textbf{UAUC Impr.}	 &  \textbf{GUC Impr.} &  \textbf{FLOPs} &  \textbf{Speed}\\ 
    \hline
    5 & 0 & 0 & 0 & 190.27M & 141.76K  \\
    10 & 0.0237\% & 0.2037\% & 0.2384\% & 194.09M & 122.60K  \\
    20 & 0.0733\% & 0.2369\% & 0.2500\% & 203.04M & 108.17K  \\
    \hline
    \end{tabular}}
    \vspace{-4pt}
\end{table}

\end{itemize}

\subsection{Visualization Study}
\begin{figure}[h]
\vspace{-0.4cm}
\centering
\includegraphics[width=.49\textwidth]{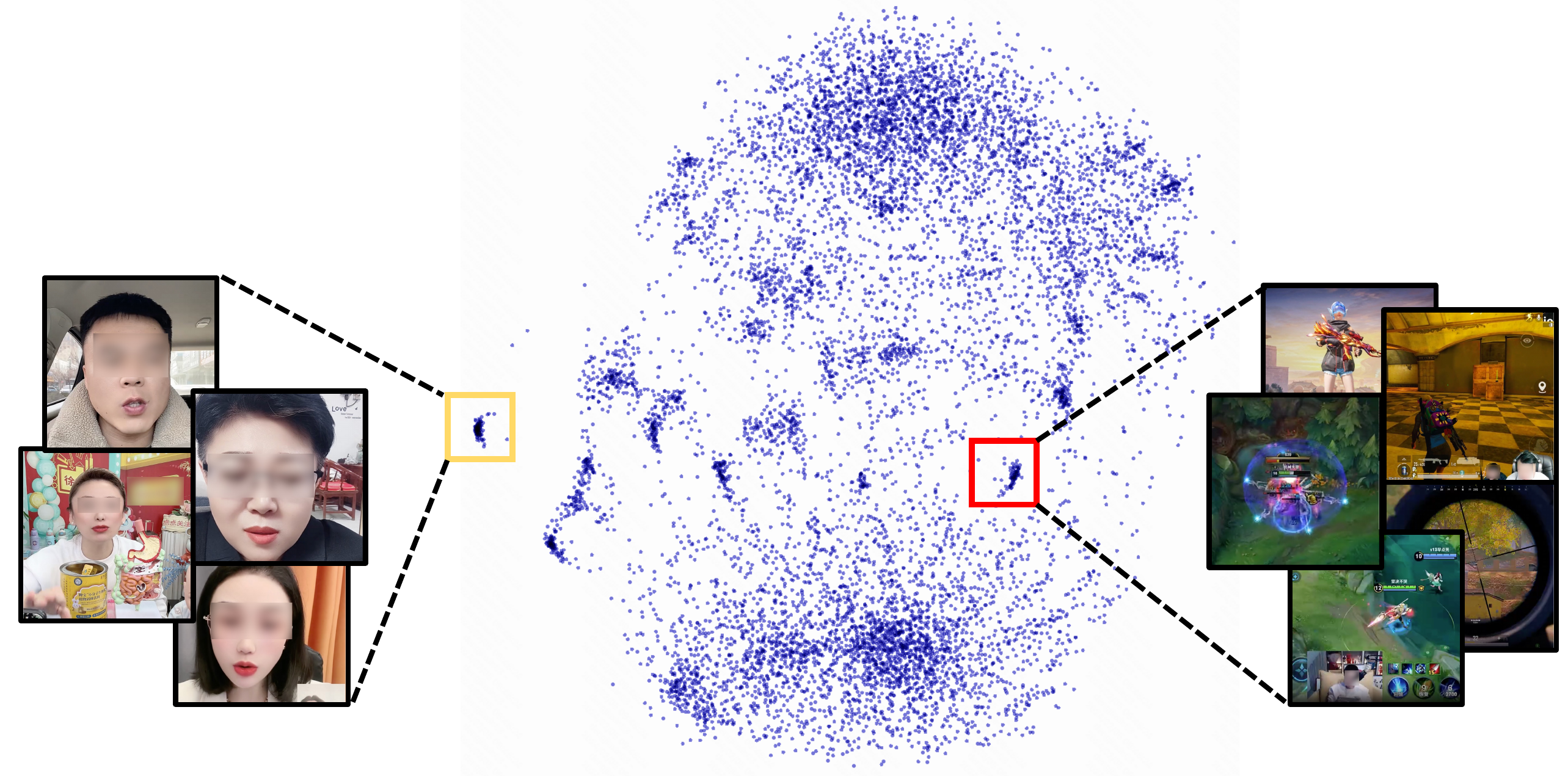}
\caption{\textbf{Visualization of the learnable query distribution in MFQ, where each point indicates an author.}}
\label{fig7}
\vspace{-0.3cm}
\end{figure}

We conduct experiment to visualize the learnable query representations in MFQ. We randomly sample 10,000 authors and visualize these representations using t-SNE \cite{van2008visualizing} in 2 dimensions, as illustrated in Figure \ref{fig7}. The points in this graph represent the sampled authors, and it is obvious that there are several distinct clustering centers and we mark two of them by the yellow and red boxes. To demonstrate the characteristics of each clustering center, we provide some visual frames for further explanation. We observe that authors in the yellow box tend to be chatting authors, while gaming authors tend to appear in the red box. These phenomena support our assumption that learnable query can represent distinctive characteristics of various types of authors.

\subsection{Study of Online Response Time}
\begin{figure}[h]
\vspace{-0.0cm}
\centering
\includegraphics[width=.49\textwidth]{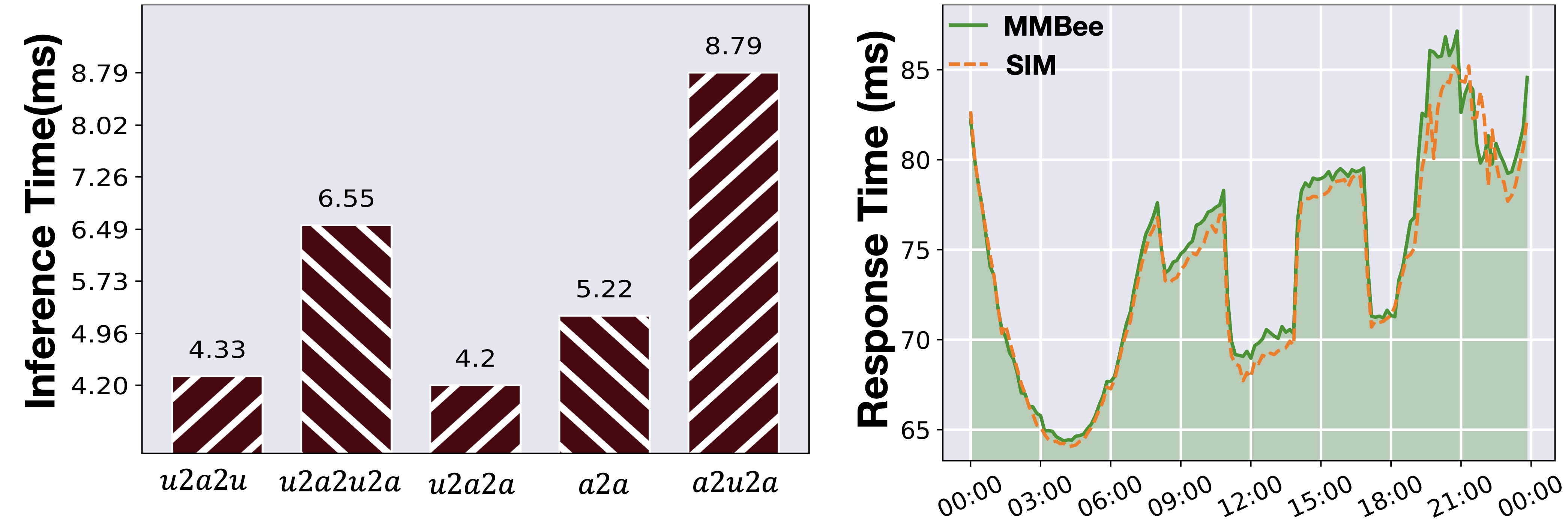}
\vspace{-0.2cm}
\caption{\textbf{Left shows the response time of different metapaths and right shows the system's overall response time change during one day.}}
\label{fig5}
\vspace{-0.4cm}
\end{figure}
We investigate the online response time when recommendation requests the KGNN server and Figure \ref{fig5} (left) shows the different response time when requesting different metapath behaviors. It is obvious that the max lag can reach 8.79 ms but this is not allowed in real-world applications. So we applied the pre-request of expansion behaviors and stored it in advance (described in Section \ref{System_Deployment}) so the online recommendation model could access the embedding server instead of walking through the graph on the fly. We evaluate the efficiency of offline storage by comparing the time cost between the baseline system and the system equipped with MMBee. The response time (in milliseconds) with millions of queries per second during Jan. 24, 2024 is presented in Figure \ref{fig5} (right), where the yellow and green lines represent the response time of the baseline system and MMBee. Empirical evidence shows that the response time of MMBee is only about 1 ms more than that of the baseline system on average, which is brought by the extra expanded graph behavior retrieving and computational overhead of inference.

\subsection{Online Result}
To evaluate the online performance of MMBee, we conduct strict online A/B tests on Kuaishou’s business scenarios of live streaming main page spanning from 2023/10/05 to 2023/10/09 and we compare the performance of MMBee and SIM with 1\% main traffic for experiments. Note that MMBee integrates our proposed MFQ and GIE into SIM backbone. We use \textbf{NGU} (Number of users who sent gifts) and \textbf{NGC} (the total number of gifts sent) as main online metrics. Online evaluation shows that MMBee has achieved \textbf{2.862\%} on NGU and \textbf{4.775\%} lift on NGC metric, which indicates that MMBee achieves much better recommendation results and brings considerable revenue increments for the platform.

\section{Conclusion}

In this paper, we propose a novel real-time multi-modal fusion and behavior expansion model called MMBee for live streaming gifting prediction. The model efficiently leverages real-time multi-modal features and effectively exploits metapath-guided expanded behaviors to enhance the performance of GTR prediction. We address two important challenges in live streaming gifting prediction, namely the multi-modal modeling and behavior sparsity, by introducing the Multi-modal Query Fusion (MFQ) and Graph-guided Interest Expansion (GIE) modules. Extensive experiments on real-world datasets demonstrate the excellent performance of MMBee.



\end{document}